\documentstyle[prl,aps,multicol,epsf]{revtex}
\begin{document} 
\draft 
\title{Optical conductivity of strongly correlated electron systems}

\author{R. Eder$^1$ , P. Wrobel$^2$, and Y. Ohta$^3$}
\address{$^1$Department of Applied and Solid State Physics, 
University of Groningen,\\ 
9747 AG Groningen, The Netherlands\\ 
$^2$ Institute for Low Temperature and Structure Research,\\
50-950 Wroclaw, Poland\\
$^3$Department of Physics, Chiba University, Chiba 263, 
Japan} 
\date{\today}
\maketitle

\begin{abstract} 
We present an exact diagonalization study of the frequency and wave 
vector dependent conductivity $\sigma(\bbox{q},\omega)$ in small clusters 
of  $2D$ $t$$-$$J$ model. Unlike the related dynamical density 
correlation function, $\sigma(\bbox{q}=0,\omega)$ in the underdoped regime
has the exchange constant $J$ as its characteristic 
energy scale and is dominated by a resonance-like excitation with frequency 
$\sim 1.7J$. We interpret this as transition to a $p$-like
excited state of a spin-bag type quasiparticle 
(or, alternatively, a tightly bound spinon-holon pair)
and show that a simple calculation based on the string picture
explains the numerical results semiquantitatively.
For doping levels $\ge 25$\% $t$ remains the only energy scale
of $\sigma(\bbox{q}=0,\omega)$.
\end{abstract} 
\pacs{74.20.-Z, 75.10.Jm, 75.50.Ee}
\begin{multicols}{2}

\begin{figure}
\epsfxsize=6.1cm
\vspace{-0.5cm}
\hspace{1cm}\epsffile{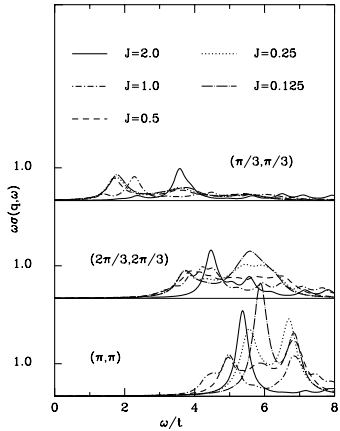}
\vspace{0ex}
\narrowtext
\caption[]{Current correlation function
$\omega\cdot\sigma(\bbox{q},\omega)$ for different momenta along the
$(1,1)$ direction in the $18$-site cluster, with two holes.
$\delta$-functions are replaced by Lorentzians of width $0.5t$}
\label{fig1} 
\end{figure}

\begin{figure}
\epsfxsize=8.2cm
\vspace{-0.5cm}
\hspace{0ex}\epsffile{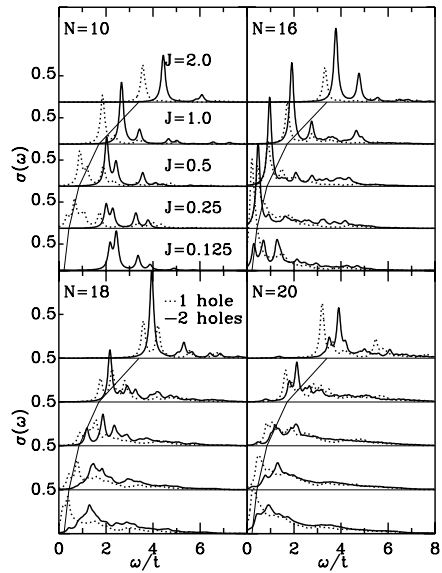}
\vspace{0cm}
\narrowtext
\caption[]{Optical conductivity $\sigma_x(\omega)$ 
for low hole concentrations in different cluster size $N$.
The thin lines indicate the points $\omega$$=$$1.7J$,
Lorentzian broadening $0.1t$.}
\label{fig2} 
\end{figure}

\begin{figure}
\epsfxsize=8.1cm
\vspace{-0.5cm}
\hspace{0.5cm}\epsffile{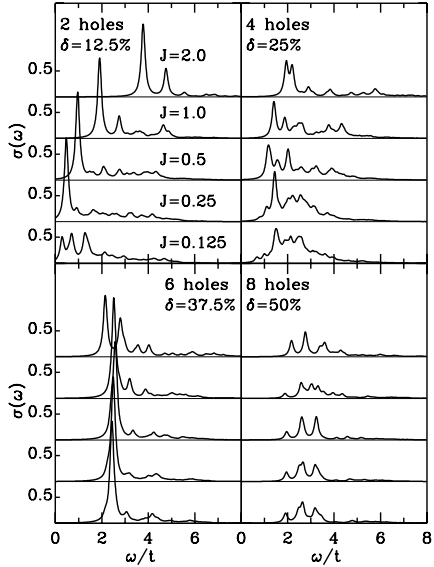}
\vspace{0cm}
\narrowtext
\caption[]{Optical conductivity $\sigma_x(\omega)$ 
for $N$$=$$16$ and different hole numbers. Lorentzian broadening $0.1t$.}
\label{fig3}
\end{figure}

\begin{figure}
\epsfxsize=10cm
\vspace{-0.5cm}
\hspace{-0.5cm}\epsffile{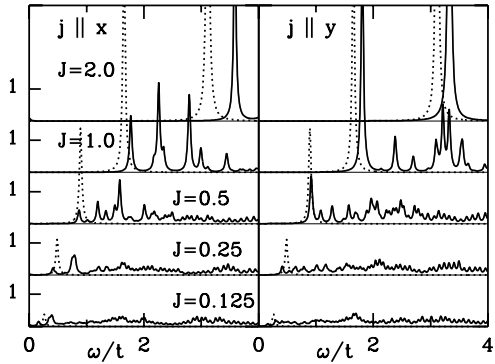}
\vspace{-4.5cm}
\narrowtext
\caption[]{Exact current correlation function
for one hole in $18$ sites (fulll line) compared to
the result of the string calculation (dotted line). 
Lorentzian broadening $0.025t$.}
\label{fig4}
\end{figure}

Electronic excitations at finite frequency
seem to be a common feature of doped Mott-Hubbard
insulators, and are observed experimentally in
the optical conductivity of cuprate superconductors\cite{uchida}, 
as well as in numerical studies of 2D strong-correlation 
models\cite{Sega,MoreoDagotto}. It is the purpose
of the present paper to present a systematic study of the optical
response function and its dependence on both, parameter values
and hole concentration, for the standard $t$$-$$J$ model.
This model is supposed to describe the low energy physics
of the $CuO_2$ planes, at least for energies which are
below the `binding energy' of a Zhang-Rice singlet;
the latter may be estimated to be $\approx 1eV$.
As a key result, the optical conductivity $\sigma(\bbox{q}=0,\omega)$
in the underdoped regime is shown to be dominated by a single excitation which
has the exchange constant $J$ as its relevant energy scale,
in contrast to e.g. the behaviour seen for the closely related
(see below) dynamical charge correlation function\cite{EderOhtaMaekawa}. 
We show that a simple calculation based on the
string picture can explain this dominant excitation as a transition to
a $p$-like excited state of a spin bag-like quasiparticle or,
alternatively, a tightly bound spinon-holon pair.
We study the finite frequency optical response,
defined as
\[
\sigma_\alpha ({\bf q},\omega)=
\sum_{\nu\neq 0} \frac{1}{\omega}\;|\langle\psi_\nu|j_\alpha({\bf q}) 
|\psi_0\rangle|^2
\; \delta( \omega - (E_\nu - E_0) ),
\]
where
$|\psi_\nu\rangle$ ($E_\nu$) denotes the $\nu^{th}$ eigenstate 
(eigenenergy)
of the system (in particluar $\nu$$=$$0$ denotes the ground state).
Also, $j_\alpha$ with $\alpha=x,y$ denotes a component of
the current operator
\[
\bbox{j}(\bbox{q}) = i 
\sum_{m,n} t_{mn}
e^{i \bbox{q}\cdot (\bbox{R}_m + \bbox{R}_n)/2 }
\;[\bbox{R}_m - \bbox{R}_n\;]
\hat{c}_{m,\sigma}^\dagger \hat{c}_{n,\sigma}.
\]
In the present study we restrict ourselves to the finite-frequency
response and disregard the `Drude peak' at $\bbox{q}$$=$$0$ and
$\omega$$=$$0$. The latter is absent in
finite clusters with periodic boundary
conditions (as used in the present study), so that its weight $D$
can only be inferred indirectly via a sum rule:
\[
\frac{ -E_{kin}}{4N} = \frac{D}{2\pi e^2} + \frac{1}{N}
\int_{0^+}^\infty d\omega \sigma_x(\bbox{q}=0,\omega).
\]
The Drude weight $D$ and its dependence
on hole doping and parameter values have been studied
previously by various authors. In 2D systems and hole concentrations
$\delta \leq 0.25$
to good approximation $D\sim \delta$, with the constant of proportionality 
being nearly independent of $J/t$\cite{DagottoDrude}. 
Since $D$ originates from the free acceleration of charge carriers in an 
applied electric field, 
this result suggests a carrier number given by the number of
doped holes, which is consistent with a variety of experiments\cite{batlogg}.
In a noninteracting single band model the Drude part
is the only contribution to $\sigma(\bbox{q}=0,\omega)$;
by contrast, the finite frequency part of $\sigma(\bbox{q}=0,\omega)$
is a special feature of correlated systems,
and the excitations which are probed by this part of the
correlation function have not yet been identified. In the present study
we focus exclusively on this feature of the
frequency dependent conductivity.\\
To begin with, Figure \ref{fig1} shows the `dispersion'
of the current correlation function
$\omega \cdot \sigma(\bbox{q},\omega)$ for different $J/t$.
The `dominant features' do not change appreciably with $J$,
i.e. the energy scale of the correlation function is $t$
(an exception is the largest peak in the
$(\pi/3,\pi/3)$ spectrum, which seems to scale with $J$ for $J$$\ge$$1$). 
This fact as well as the dispersion of the spectral weight,
which resembles a `smeared out' free electron band
of width $\sim 8t$, is to be expected on the grounds of
analogous results for the dynamical charge 
correlation function (DCF)\cite{EderOhtaMaekawa}.
For e.g. $\bbox{q}$ parallel to $(1,0)$ one has
$[H, n_{\bbox{q}}]= 2 \sin(q_x/2) j_x(\bbox{q})$, so that the
current correlation for finite momentum transfer $\bbox{q}$
is closely related to the DCF, and in principle provides no new
information. This is not the case, however, for
the optical conductivity, $\sigma(\bbox{q}=0,\omega)$, so that we
henceforth consider only case $\bbox{q}$$=$$0$ and drop
the wave vector $\bbox{q}$ for simplicity.\\
\noindent
Then, $\sigma(\omega)$ is shown in Figure \ref{fig2}
for $1$ and $2$ holes and different $J$, for all
available cluster sizes $N$ (except for $1$ hole in
$10$ sites, $J=0.125$, which already is in the fully
polarized Nagaoka state).
Comparing with the spectra in Figure \ref{fig1} 
which correspond to large momentum transfer,
such as $(\pi,\pi)$, some substantial differences are obvious:
for a single hole as well as for two holes and
larger values of $J$ the most prominent feature is a pronounced 
maximum of intensity at the
lower edge of the spectra, whose energy scales with $J$.
This maximum seems to have a more resonance-like character
(i.e. a substantial `broadening') for
two holes and smaller $J$; even in this case, however,
the spectra do have
a $J$-dependent offset which is nearly the same as for
the single hole. 
The $16$-site cluster is
exceptional, in that the peak 
still is very pronounced even for two holes, which may be an effect
of its special geometry.
As mentioned above, for the smallest momentum transfer, $(\pi/3,\pi/3)$,
and large $J$
a similar scaling with $J$ can be seen also in Figure \ref{fig1}.
Since $\sigma(\bbox{q},\omega)$ ultimately must be a continuous function
of $\bbox{q}$, there has to be a crossover of energy scales
from small to large $\bbox{q}$. The wave vector where this crossover occurs
apparently depends on $J$ and probably is smaller than
$(\pi/3,\pi/3)$ for $J$$\le$$0.5$.\\
Next, the evolution of the optical conductivity with doping is
shown in Figure \ref{fig3}. There is a rather obvious
crossover between the low doping region with a $J$ dependent
offset, and the
higher doping region where $t$ is the only remaining
energy scale.
For `physical' values of $J$,
i.e. $J\approx 0.4$, the two-hole spectra in
$16$, $18$ and $20$ sites(corresponding to $\delta$$=$$0.1-0.125$)
still show the $J$-dependent offset, wheras already at $\delta$$=$$0.2$
($2$ holes in $10$ sites) no more more scaling with $J$ can be seen,
so that the critical concentration $\delta_c$ falls between these two
values: $0.125$$<$$\delta_c$$<$$0.2$. Similarly one may estimate that
$0.2$$<$$\delta_c$$<0.25$ for $J$$=$$1$, and $\delta_c$$\approx$$0.25$
for $J$$=$$2$. It is tempting to speculate that the
crossover for physical values is close to the `optimal doping',
in the the cuprates, $\delta$$=$$0.15$ but it should be noted
that exact diagonalization
allows to vary the hole concentration only in rather coarse steps,
so that it is difficult to give a really accurate estimate for $\delta_c$.
We note that such a crossover is also quite consistent with
analogous results for spin and charge correlation 
functions\cite{EderOhtaMaekawa,EderOhta}. Moreover
the doping dependence of the
\noindent
Drude weight changes from the 
strong-correlation form
$D \propto \delta$ to essentially free electron-like behaviour
at approximately the same hole concentration\cite{DagottoDrude}.
All of this indicates that a profound reconstruction of the electronic
structure takes place at this concentration.\\ 
We focus on the physically more interesting low doping region and proceed to
an interpretation of the dominant `$J$-peak' in $\sigma(\omega)$.
The similarity between the single and two hole cases suggests that
the essential physics is already present for a single hole
in an antiferromagnet, a limiting case that is reasonably
well understood in terms of the `string' picture.
Following Ref. \cite{EderOhtaMaekawa} we assume that
`charge spectra' in the low doping region
are dominated by excitations of internal degrees 
of freedom of the spin bag like quasiparticles.
More precisely, we assume that the quasiparticles in the
low doping regime correspond to a hole oscillating rapidly 
(i.e. on an energy scale $\sim t$)
within a region of enhanced 
spin disorder\cite{Schrieffer,EderBecker,GanHedegard}.
The entire quasiparticle (i.e. `bare hole' plus `defect region') 
then drifts slowly (i.e. on an energy scale $J$) and coherently
through the system. 
\noindent
A different way to express this
picture would be `incomplete spin charge separation', where the 
`holon' (=bare hole) and the `spinon' (=first spin defect created by the 
hopping hole) remain tied together firmly by a `string', resulting
in a bound state.
As discussed in Ref. \cite{EderOhtaMaekawa} both, the different 
energy scales of spin and charge 
correlation function and their different scaling behaviour with hole number
$n_h$ can be understood immediately by assuming that the density operator
couples only to the rapid incoherent motion, whereas the
spin operator couples only to the slow coherent motion.
Extending this line of thought
we would expect that the current operator, which acts
on charge degrees of freedom as well,
should also couple to the rapid incoherent motion, 
i.e. excite `internal degrees of freedom' of the spin bags.
Due to the high symmetry for 
$\bbox{q}=0$, however, there is predominantly one special type of state
which is excited by the current operator:
we may assume, that the drifting spin bag 
(or,phrased differently, the bound spinon-holon pair) to good 
approximation has 
maximum point group symmetry, i.e. it is an $s$-like state.
The current operator for $\bbox{q}=0$ is a vector, and hence must 
obey a `dipole selection rule', in other words: it can couple
only to `$p$-like' excited states of the spin bag (or spinon-holon pair).\\
We make these considerations more quantitative,
using the formalism of Ref. \cite{EderBecker}.
We begin with a hole created at some site $i$ 
in the Neel state. Acting with the
hopping term, we obtain `string states', where the hole is
connected to its `starting point' $i$ 
by a trace of misalligned spins\cite{Bulaevskii}.
We denote a state with $\nu$ such defects as $|i,\nu, {\cal P}\rangle$
where ${\cal P}$ is shorthand for a set of numbers which parameterize
the geometry of the hole path. The wave function
for a hole which is trapped in
the string potential around the site $i$ then reads
\begin{equation}
|\Psi_{i,\lambda} \rangle = \sum_\nu \alpha_{\nu,\lambda} \sum_{ \cal P}
\phi_{\lambda}( {\cal P}) 
|i,\nu, {\cal P}\rangle.
\label{local}
\end{equation}
Here, $\phi( {\cal P})$ denotes an extra phase factor,
which determines the `orbital symmetry' $\lambda$ of the self-trapped state.
Let $\bbox{e}$ denote a unit vector in the direction of the first step
of the path ${\cal P}$. Then,
$\phi_\lambda( {\cal P})=1$ will give a totally symmetric
(`$s$-like') spin bag, $\phi_\lambda( {\cal P})=e_x$  will give
a state with $p_x$-type symmetry etc.
The coefficients $\alpha_{\nu,\lambda}$, which play the
role of a `radial wave function' associated with the
orbital symmetry $\lambda$, are assumed to
depend only on the length of the string, $\nu$.
If we assume that the magnetic frustration in the system
increases linearly with the string length $\nu$\cite{Bulaevskii},
and introduce 
$\beta_{\mu,\lambda}=\alpha_{\mu,\lambda} (z-1)^{\mu/2}$
(with $z$ the coordination number), the latter function
can be determined from the (numerical) solution of the 
`radial Schr\"odinger equation'\cite{EderBecker}
\begin{eqnarray}
- \frac{z}{z-1}\tilde{t} \beta_{1,\lambda}+ 
2J \beta_{0,\lambda} &=& E_\lambda\beta_{0,\lambda}
\nonumber \\ 
-\tilde{t}( \;\beta_{\nu+1,\lambda} + \beta_{\nu-1,\lambda}\;) +
J(\frac{5}{2}+\nu)\beta_{\nu,\lambda} &=& E_\lambda \beta_{\nu,\lambda},
\label{ssb}
\end{eqnarray}
where $\tilde{t}=\sqrt{z-1} t$. The
normalization condition reads
\[
\beta_{0,\lambda}^2 + \frac{z}{n_\lambda(z-1)} 
\sum_\nu \beta_{\nu,\lambda}^2 =1,
\]
with $n_s=1$, $n_p=2$ and
$\beta_{0,p}=0$.
(and consequently the first equation (\ref{ssb}) being omitted for
$\lambda=p$). Next,
a propagating spin bag with orbital symmetry $\lambda$ and momentum
$\bbox{k}_0$ would be described by
\begin{equation}
| \Psi_\lambda(\bbox{k}_0)\rangle = 
\sqrt{\frac{2}{N}} \sum_j e^{i \bbox{k}_0 \cdot \bbox{R}_j }
|\Psi_{j,\lambda}\rangle.
\label{coh}
\end{equation}
When comparing with the numerical results we choose
the $18$-site cluster, where the single hole ground state
momentum is $\bbox{k}_0$$=$$(2\pi/3,0)$
(this avoids complications due to the spurious degeneracies
in the $4\times 4$ cluster).
The little group of $\bbox{k}_0$
comprises $E$ and the reflection by the $(1,0)$ direction,
the ground state wave function being even.
Choosing the current in $(1,0)$ direction the final states
must be even as well and thus should correspond
to a propagating $p_x$-like spin bag, whereas 
choosing the current in $(0,1)$ should couple to a propagating
$p_y$-like spin bag.
Assuming that the dispersion of the states (\ref{coh}) is solely due to
the relaxation of the strings through the transverse part
of the Heisenberg exchange and 
following Ref. \cite{EderBecker} in the computation of
the respective matrix elements, we obtain the
dispersion of the $s$-like state as
$E_s(\bbox{k}) = E_s + 4 h_s ( [\cos(k_x)+\cos(k_y)]^2-1)$, 
and that of a $p_\alpha$ state as
$E_{p}(\bbox{k})  = E_p + 2 h_p cos(2 k_\alpha)$, with
 the `spin-flip matrix element'
$h_\lambda = \frac{J}{(z-1)^{n_\lambda}}
\sum_{\nu=0}^\infty \beta_{\nu,\lambda} \beta_{\nu+2,\lambda}$.
The latter describes the truncation of the string by the Heisenberg
exchange. 
Differences between those energies at $\bbox{k}_0$ 
give the excitation energies in the optical conductivity.
The matrix element of the current operator is
$\langle \Psi_s(\bbox{k}_0)
| j_\alpha |\Psi_{p_\alpha}(\bbox{k}_0) \rangle$$=$$-2it
 \sum_{\nu=0}^\infty
(\alpha_{\nu,s}\alpha_{\nu+1,p} -
\alpha_{\nu,p}\alpha_{\nu+1,s})$.
Then, Figure \ref{fig4} compares the current correlation function
obtained by Lanczos diagonalization with the result of the
string calculation for transitions to the lowest $p$-like state.
We note that the string calculation correctly reproduces the
$J$-dependence of the offset. For larger $J$ and current perpendicular
to the momentum, even the weight of the dominant peak is reproduced quite
well. On the other hand, for smaller $J$ the `peak' at the bottom of the 
Lanczos spectra seems to `decay' into a multitude of small peaks, i.e.
there is a strong damping. This is to be expected,
because the $p$-like state first of all is high in energy
(on the energy scale of spin fluctuations) and, having a node in the center,
necessarily has a large spatial extent; most probably it is therefore very
susceptible to scattering from spin excitations and/or
finite size effects, so that the $p$-like
`state' rather has the character of a `resonance'.
While it seems that the string calculation reproduces the rapid decrease of
spectral weight, any more quantitative comparison must fail due to this
decay of the peak. Taking into account its highly oversimplified nature,
the string calculation nevertheless predicts some trends reasonably well,
and we believe that it actually is the correct explanation for the
observed behaviour. It should be noted that a calculation for
`spinon'-`holon' pair, which is bound by a potential $\sim J$ times their 
distance would almost inevitably give an identical result.
Finally, the apparent scaling of the dominant peak
in the current-correlation function
for $\bbox{q}$$=$$(\pi/3,\pi/3)$ with $J$, which seems to
occur for larger $J$, may be an indication that a propagating
$p$-like spin bag represents an approximate eigenstate
also for small finite momentua.\\
In summary, we have presented a systematical study of the
optical conductivity in the 2D $t$$-$$J$ model. We found that over the
entire `underdoped' range of hole concentrations, the optical spectrum
is dominated by a single excitation at an energy of $\approx 1.7J$.
The appearance of the exchange constant
$J$ as the energy scale of a `charge spectrum' is not consistent with an
interpretation in terms of spin charge separation, where the
charge spectra would be described by the excitations of holons,
whose energy scale is $t$; the latter would be neccessary 
to be consistent with the finite momentum current correlation and
density correlation functions.
On the other hand, an interpretation in terms of the `string' 
picture, (or alternatively the assumption of tightly bound
spinon-holon pair)
where the dominant excitation in $\sigma(\omega)$
is interpreted as transition to the lowest excited state of the `spin bag' 
with $p$-type symmetry, readily provides a semiquantitative explanation 
for the numerical data. The absence of an appreciable
finite frequency component in $1D$ then fits nicely with this picture:
in $1D$ spinon and holon truly unbind, so that excited `bound states'
which could serve as final states in $\sigma(\omega)$ do not exist.\\
We note that `incomplete spin-charge separation', i.e. the formation
of tightly bound spinon-holon pairs, in reality would imply absence of 
spin charge
separation. When probed with wave lengths larger than its spatial extent,
or very low energies, 
the spinon-holon pair should behave just like a spin-1/2 Fermion, so that
the `effective' low energy theory would be a Fermi liquid of
quasiparticles corresponding to the doped holes.
This scenario is indeed suggested by diagonalization\cite{rigid,jaklic}
and other\cite{Lee,Dagottoflat} studies. The nature of the
`transition' at higher doping levels, and the resulting
ground state remains to be clarified.\\
Financial support of R. E. by the European Community and of P. W.
by KBN is most gratefully acknowledged.
 
\end{multicols}
\end{document}